\documentclass[12pt,preprint]{aastex}

\begin{document}

\title{Observed and Physical Properties of Core-Collapse Supernovae}
\author{Mario Hamuy\altaffilmark{1}}
\affil{The Observatories of the Carnegie Institution of Washington, 813 Santa Barbara Street, Pasadena, CA 91101}
\email{mhamuy@ociw.edu}

\altaffiltext{1}{Hubble Fellow}

\begin{abstract}

I use photometry and spectroscopy data for 24 Type II plateau supernovae to
examine their observed and physical properties. This dataset shows that
these objects encompass a wide range of $\sim$5 mag in their plateau luminosities,
their expansion velocities vary by $\times$5, and the nickel masses produced in these explosions
go from 0.0016 to 0.26 $M_\odot$.  From a subset of 16 objects I find that the explosion
energies vary between 0.6$\times$ and 5.5$\times$10$^{51}$ ergs, the ejected masses encompass
the range 14-56 $M_\odot$, and the progenitors' radii go from 80 to 600 $R_\odot$.
Despite this great diversity several regularities emerge,
which reveal that there is a continuum in the properties of these objects
from the faint, low-energy, nickel-poor SNe~1997D and 1999br, to the bright,
high-energy, nickel-rich SN~1992am. This study provides evidence that more
massive progenitors produce more energetic explosions, thus suggesting that
the outcome of the core collapse is somewhat determined by the envelope mass.
I find also that supernovae with greater energies produce more nickel.
Similar relationships appear to hold for Type Ib/c supernovae, which suggests that
both Type II and Type Ib/c supernovae share the same core physics. When the whole
sample of core collapse objects is considered, there is a continous distribution of
energies below 8$\times$10$^{51}$ ergs. Far above in energy scale and nickel production
lies the extreme hypernova 1998bw, the only supernova firmly associated to a GRB.

\end{abstract}


\section{INTRODUCTION}

The advent of new telescopes and better detectors is causing a rapid increase in
the quality and quantity of observations obtained for supernovae (SNe, hereafter)
of all types. Although the field of Type Ia SNe (exploding white dwarfs) has developed
considerably faster in recent years (due to the widely acknowledged importance of
such objects as cosmological probes), there is a growing body of data for core
collapse SNe. In this paper I collect all of the available data on hydrogen-rich
plateau Type II SNe (those undergoing little interaction with the circumstellar medium,
SNe~II-P hereafter), with the purpose to better understand the nature of such objects.

I start in section \ref{OM} by summarizing the observational material available on
24 SNe~II-P, after which (Sec. \ref{OP}) I proceed to examine their great diversity
and the correlations among the observed parameters. 
Using the hydrodynamic models of \citet[hereafter LN83, LN85]{litvinova83,litvinova85} I go a step
further and derive physical parameters (explosion energies, progenitor masses and radii)
for 13 SNe~II-P (Sec. \ref{PP}). Although the statistics are still poor, this study shows that
progenitors with greater masses produce more energetic explosions and synthesize more nickel.
These correlations provide valuable clues and a better insight on the explosion mechanisms.
In section \ref{PCCS} I combine the physical parameters of the SNe~II-P with
those previously published for SNe~Ib/c. It appears that all core collapse SNe display 
the same correlations, which suggests that all of these objects share the same core physics.
I discuss the properties of all core collapse SNe and how hypernovae fit in this group.

\section{OBSERVATIONAL MATERIAL}
\label{OM}

Table \ref{SNII_tab} lists the 24 SNe~II-P for which I have
photometric and spectroscopic data. For each SN this table includes the
heliocentric redshift (from the NASA/IPAC Extragalactic Database or my own measurement),
reddening due to our own Galaxy \citep{schlegel98}, host galaxy
extinction, the distance, and the method used to derive the distance.

In two cases I use Cepheid distances in the scale published by \citet[F00]{ferrarese00}.
For five objects it is possible to assign the SN host galaxy to a galaxy group with
surface brightness fluctuation (SBF) distances \citep{tonry01} in the F00 Cepheid scale
(adopting an uncertainty of 1 Mpc to account for cluster depth). For the 9 SNe
which are not sufficiently far in the Hubble flow ($cz$$<$3000 km~s$^{-1}$) and do not have
SBF or Cepheid distances, it proves necessary to correct their observed redshifts
in order to account for peculiar motions of their host galaxies. For this purpose
I adopt the parametric model for peculiar flows of \citet{tonry00} which includes
infall into Virgo and the Great Attractors, an overall dipole, and a cosmic thermal
velocity dispersion of 187 km s$^{-1}$. Given the observed Cosmic Microwave Background (CMB)
redshift the model yields a SBF distance in the F00 scale.
For the 8 most distant objects with CMB redshifts greater than 3000 km~s$^{-1}$ I use
their redshifts to compute the distances and an associated velocity dispersion of 187 km s$^{-1}$.
To be consistent with the method employed for the nearby SNe, I adopt the best value for the Hubble
constant in the F00 scale, namely, $H_0$=68$\pm$2 from SNe~Ia \citep{gibson00}. Note however that the
SBF distances in the F00 scale yield $H_0$=77$\pm$4 \citep{tonry00}, which suggests that the SBF and SN~Ia distances
could be systematically different (for a different view see \citet{ajhar01}, who claim
that the SBF and SNe~Ia distances agree very well). For now I prefer to adopt $H_0$=68 since
this value is determined from SNe~Ia well in the quiet Hubble flow (unlike the value
derived from SBF). Certainly, it would be more convenient to use the new Cepheid scale
reported by the Hubble Space Telescope (HST) Key Project \citep{freedman01} instead of the F00 scale
since the new scale reconciles the SBF and SNe~Ia methods, but it will necessary to wait
until the parametric model for peculiar flows of \citet{tonry00} is updated.

The estimate of the amount of foreground visual extinction is under good control ($\sigma$=0.06 mag)
thanks to the IR dust maps of \citet{schlegel98}. The determination of absorption in
the host galaxy, on the other hand, is more challenging. Since SNe~II occur near HII regions,
this is potentially a significant problem. To zero order SNe~II-P should all reach the same
temperature of hydrogen recombination during the plateau phase, so a measurement of the
color should give directly the color excess due to dust absorption. Unfortunately, significant variations
between 6,000-12,000 K are expected for the photosphere depending on the H/He abundance ratio
\citep{arnett96} which limits the precision of the method to estimate color excesses.
Keeping this caveat in mind, I proceed to use the observed colors to estimate $A_{host}(V)$
assuming that all SNe reach the same color at the end of the plateau. For this
purpose I adopt the well-studied SN~1999em as the reference for the intrinsic color, and
the $A_{host}(V)$=0.18 value derived by \citet{baron00} from detailed theoretical modeling of
the spectra of SN~1999em. As Table \ref{SNII_tab} shows it, for 22 SNe it is possible to
use their $B-V$ colors to derive extinction. A concerning problem with the $B-V$
method is that it yields negative reddenings for 10 SNe. This is particularly  pronounced
among the historical SNe, reaching $A_{host}(V)$=-1.2 for SN~1970G. It is possible that part
of the problem is due to inadequate transformations of the photographic magnitudes into the
standard Johnson system, or to background contamination by the host galaxy. However, even
SN~1999cr (with modern CCD photometry) yields a negative value of $A_{host}(V)$=-0.75, which
is well beyond the photometric errors. Perhaps this could be due to metallicity
effects which are expected to be stronger in the $B$ band where line blanketing
is stronger. For 17 SNe I use their $V-I$ colors to derive an independent reddening estimate.
This method is much well behaved: only SN~1992af yields a modest negative reddening of $A_{host}(V)$=-0.2.
Ideally it would be more convenient to use the $V-I$ extinction values -- which are expected to
be less sensitive to metallicity effects -- but, since I do not have $VI$ photometry for
all SNe, in what follows I simply use the average of the $B-V$ and $V-I$ extinction values
or the single-color value when only one color is available. I can estimate the uncertainties
in $A_{host}(V)$ by comparing the difference in reddening yielded by both methods.
Such differences amount to 0.46 mag on average, which implies a minimum error of 0.23 mag
in the reddening estimate from an individual color. To be conservative I assume $\pm$0.3 mag
in $A_{host}(V)$ for all SNe.

\section{OBSERVED PARAMETERS FOR TYPE II SUPERNOVAE} 
\label{OP}

Table \ref{50t_tab} summarizes some observables that can be measured for the 24 SNe~II including:
the time of explosion ($t_0$) which comes from a Baade-Wesselink analysis and/or considerations
about the discovery and pre-discovery image epochs;
the observed $V$ magnitude near the middle of the plateau ($V_{50}$);
the corresponding absolute $V$ magnitude corrected for extinction ($M^V_{50}$);
the SN ejecta velocity near the middle of the plateau ($v_{50}$) measured from
the minimum of the Fe II~$\lambda$5169 line (corrected for host galaxy redshift)
with an adopted uncertainty of 300 km s$^{-1}$ for all SNe;
the fiducial time ($t_{50}$) at which I measure $V_{50}$ and $v_{50}$ (arbitrarily chosen to be 50
rest-frame days after explosion);
the characteristic $V$ magnitude of the exponential tail ($V_t$);
the time ($t_t$) at which I measure $V_t$;
the nickel mass ($M_{Ni}$) produced in the explosion;
and the specific data sources of photometry and spectroscopy.

\begin{figure}
\epsscale{1.0}
\plotone{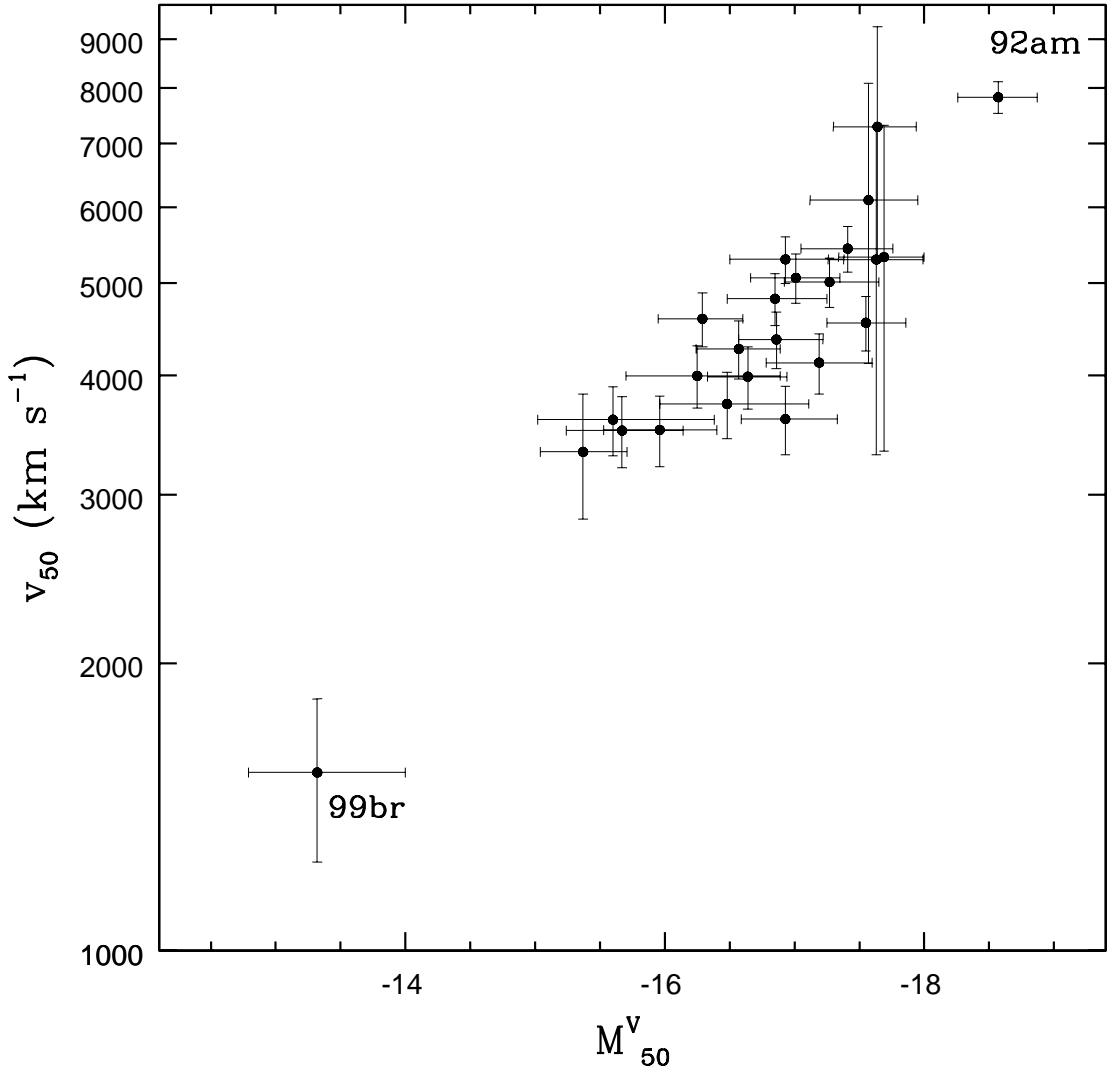}
\caption{Expansion velocities from Fe II $\lambda$5169 versus absolute $V$ magnitude,
both measured in the middle of the plateau (day 50) of 24 SNe~II-P.
\label{L_v_fig}}
\end{figure}

\begin{figure}
\epsscale{1.0}
\plotone{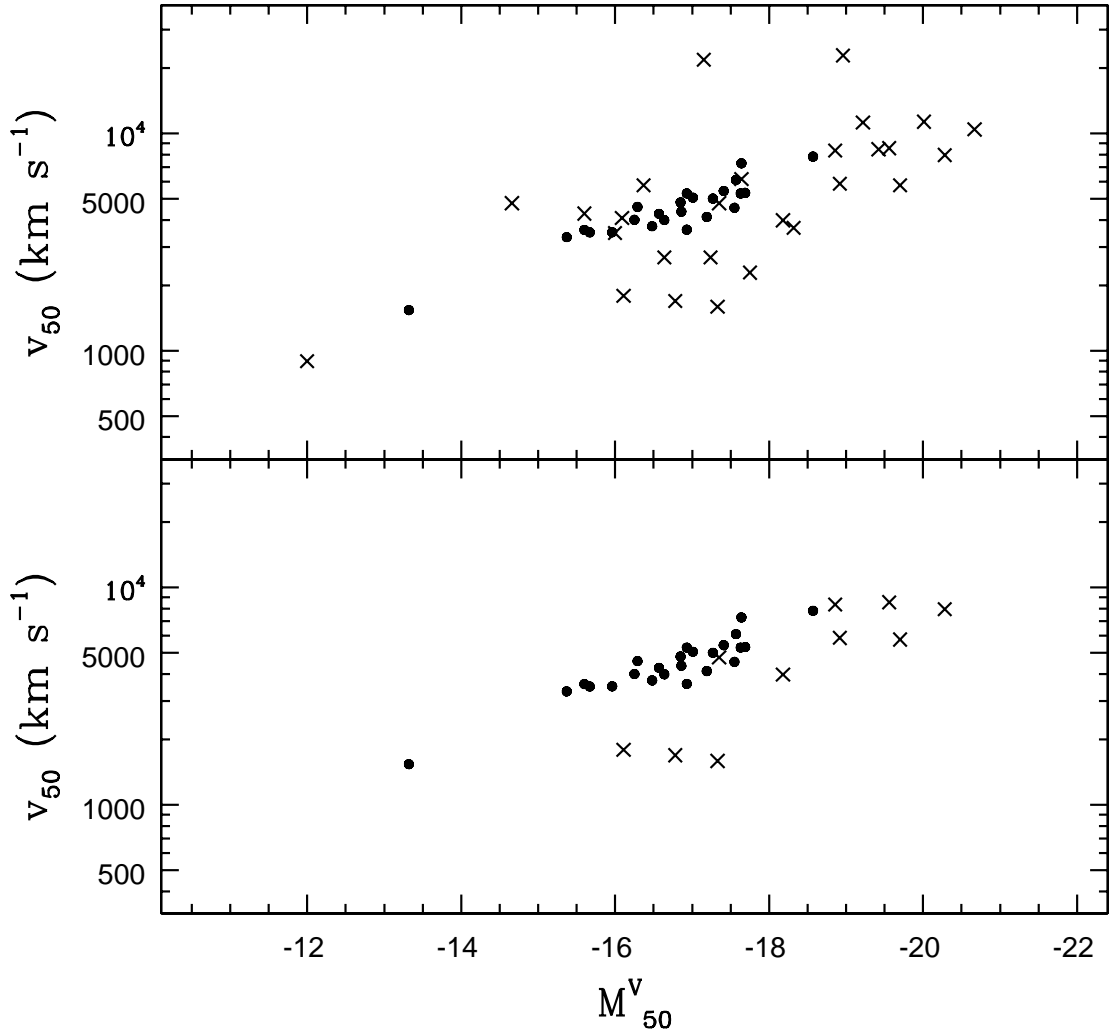}
\caption{ (top) Luminosity-velocity relation for 24 SNe~II-P (solid points)
and the 27 models of LN83 and LN85 (crosses). (bottom) Same as above, but
for a restricted set of models with $\geq$ 8 $M_\odot$.
\label{L_v_1_fig}}
\end{figure}

Figure \ref{L_v_fig} shows a comparison between the SN plateau luminosities
and their expansion velocities. The correlation between $M^V_{50}$ and $v_{50}$ 
is quite evident and proves similar to that previously reported by \citet{hamuy02}
from a smaller sample of SNe~II. This result reflects the fact that, while
the explosion energy increases so do the kinetic and internal energies.
This correlation implies that the SN luminosities can be standardized
to a level of $\sim$0.3 mag from a spectroscopic measurement of the SN ejecta velocity.
This method (named Standardized Candle Method, or SCM for short) suggests that SNe~II-P have a potential utility
as cosmological probes. A comparison of this empirical correlation to
the 27 models of LN83 and LN85 can be seen in the top panel of Figure \ref{L_v_1_fig}.
Although there is reasonable agreement between observations and theory,
the models (represented with crosses) show substantially greater scatter
than the observed quantities. It must be pointed out, however, that several of
the LN83 and LN85 calculations are for progenitors with less than 4 $M_\odot$,
which seems unrealistically low (see section \ref{PP}). When the sample of models is restricted to
a more realistic subset of progenitor masses ($\geq$ 8 $M_\odot$)
the agreement is significantly better, as can be appreciated in the bottom panel
of Figure \ref{L_v_1_fig}. It is clear that the luminosity-velocity relation
is also present in the theoretical calculations, although it seems that nature
produces a narrower correlation.

The nickel masses ($M_{Ni}$) listed in Table \ref{50t_tab} are derived from the brightness of the SN exponential tails,
assuming that all of the $\gamma$ rays due to $^{56}$Co $\rightarrow$ $^{56}$Fe are fully thermalized
($^{56}$Co is the daughter of $^{56}$Ni, which has a half life of only 6.1 days).
This is a reasonable assumption given that most, if not all, SNe~II-P have
late-time decline rates consistent with $^{56}$Co $\rightarrow$ $^{56}$Fe.
The first step in this calculation is the conversion of $V_t$
into a bolometric luminosity which can be accomplished with the following formula,

\begin{equation}
log_{10}L_t~=~\frac{-[V_t~-~A_{GAL}(V)~-~A_{host}(V)~+~BC]~+~5~log_{10}D~-8.14} {2.5},
\label{bollumeqn}
\end{equation}

\noindent where $L_t$ is the tail luminosity (ergs s$^{-1}$), $D$ is the distance in cm, $BC$ is a bolometric correction that
permits one to transform $V$ magnitudes into bolometric magnitudes, and the additive
constant provides the conversion from Vega magnitudes into cgs units. From SN~1987A
and SN~1999em I found that $BC$=0.26$\pm$0.06 during the nebular phase \citep{hamuy01a}.
Once the tail luminosity is computed the nickel mass can be found via,

\begin{equation}
M_{Ni}~=7.866\times10^{-44}\times L_t\times~exp \left\{ \frac {(t_t - t_0)/(1+z) - 6.1} {111.26} \right\} M_\odot,
\label{nieqn}
\end{equation}

\noindent where 6.1 is the half-life (in days) of $^{56}$Ni 
and 111.26 is the e-folding time (in days) of the $^{56}$Co decay, 
each of which releases 3.57 MeV in the form of $\gamma$ rays \citep{woosley89}.
The nickel masses resulting from this method are given in Table \ref{50t_tab} for 20 SNe,
along with values independently derived for SN~1987A and SN~1997D
by \citet{arnett96} and \citet{zampieri02}, respectively. The nickel masses of this sample
show a remarkably wide range: while SN~1999br yielded only 0.0016 $M_\odot$, SN~1992am
produced 0.26 $M_\odot$ of $^{56}$Ni (in good agreement with the 0.3 $M_\odot$ value previously
reported by \citet{schmidt94b}).
This result is clearly inconsistent with previous claims that SNe~II-P produce nearly the same
amount of nickel \citep{hamuy90,patat94}, but in good agreement with more recent studies \citep{turatto98,sollerman02}.
Next I proceed to examine how $M_{Ni}$ is related to the other observables.

\begin{figure}
\epsscale{1.0}
\plotone{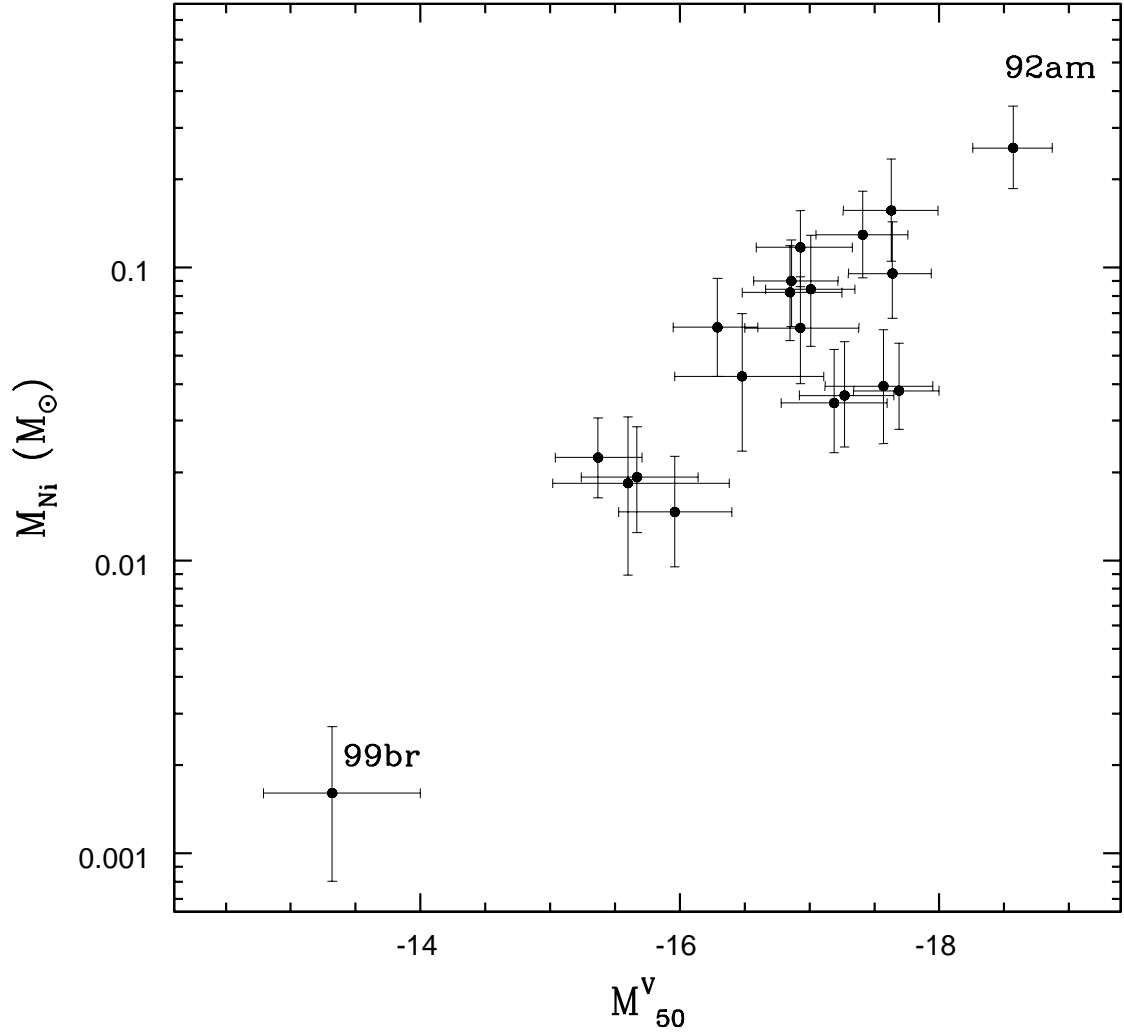}
\caption{Mass of freshly synthesized $^{56}$Ni versus
plateau luminosity measured 50 days after explosion.
\label{L_Ni_fig}}
\end{figure}

\begin{figure}
\epsscale{1.0}
\plotone{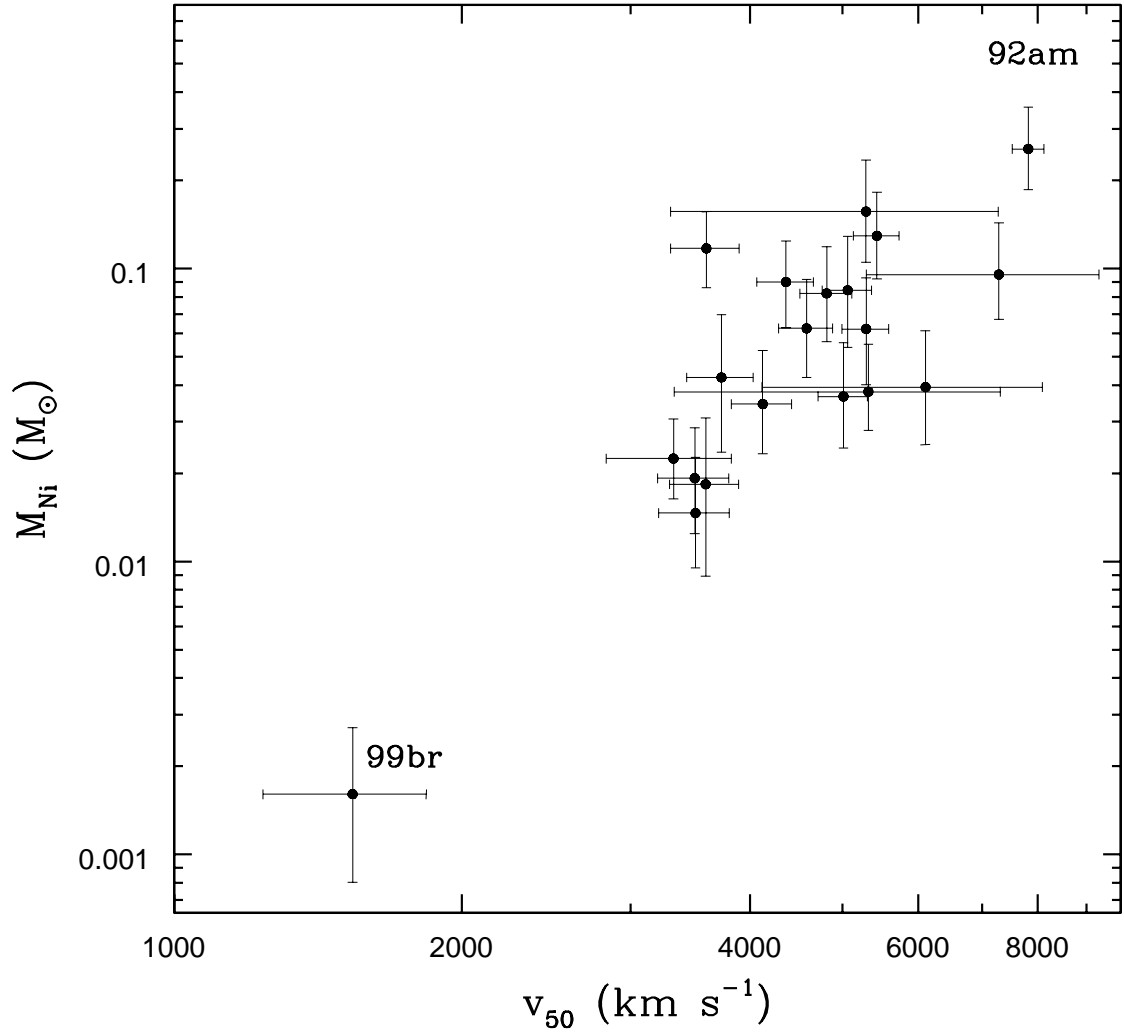}
\caption{Mass of freshly synthesized $^{56}$Ni versus
expansion velocities measured from Fe II $\lambda$5169 in the middle of the plateau (day 50).
\label{v_Ni_fig}}
\end{figure}

In numerical simulations the shock wave generated by the collapse
of the core propagates through the star's envelope, heating the material
and triggering nuclear processing in the layers above the core where
the temperatures are sufficiently high. Since the internal temperature
is determined by the progenitor's radius ($R_0$) and the explosion
energy ($E$) by the relation

\begin{equation}
T \approx \left[ \frac {3E} {4 \pi R_0^3 a} \right]^{1/4},
\label{tempeqn}
\end{equation}

\noindent the degree of nucleosynthesis is expected to be relatively greater
for SNe with smaller progenitors and greater energies \citep{weaver80}, at
least to zero order. In reality the physics is more complex, and
the amount of observed nickel depends also on how much of the material
located at the bottom of the envelope falls back to the newborn neutron
star (or black hole). Theory as yet provides no physical constraints to
this process and the amount of infalling material is freely adjusted via a
``mass-cut'' parameter. Since models currently offer no predictions
on how the explosion parameters affect the degree of nucleosynthesis in core collapse
SNe, observations can play an important role to placing constraints on the explosion mechanisms.
This issue can be examined by comparing how the nickel mass produced in the explosion is
related to the SN plateau properties (velocity and luminosity) which are determined by the
explosion parameters \citep[LN83,~LN85]{arnett96,popov93}. Among the objects of the sample,
SN~1992am is the one with the greatest nickel yield (0.26 $M_\odot$) and the brightest plateau ($M^V_{50}$=-18.57).
On the other end, SN~1999br is characterized by a dim plateau ($M^V_{50}$=-13.32) and 
a small Ni production of only 0.0016 $M_\odot$.
This pair of objects suggests that the plateau luminosity is correlated
with the nickel mass. To examine this issue, Figure \ref{L_Ni_fig} shows
$M^V_{50}$ versus $M_{Ni}$. There is clear evidence that SNe with brighter plateaus produce
more nickel. Since the plateau luminosities and velocities are tightly correlated
(Fig. \ref{L_v_fig}), it is expected that $v_{50}$ is correlated with $M_{Ni}$.
This is the case, indeed, as can be seen in Figure \ref{v_Ni_fig}, where 
SN~1992am and SN~1999br again appear as extreme objects. Since the kinetic energy 
comprises 90\% of the explosion energy of SNe~II \citep{arnett96}, this result suggests
that SNe with greater explosion energies undergo more nuclear burning. It must be
kept in kind, however, that not all SNe~II-P necessarily eject the same mass, so it is possible that
their expansion velocities do not provide a direct measure of their kinetic energies.
In the next section I examine this point in more detail.

\section{PHYSICAL PARAMETERS FOR TYPE II SUPERNOVAE} 
\label{PP}

In an elegant paper \citet{arnett80} derived analytic solutions for lightcurves of
SNe~II-P with the purpose to derive physical parameters for such objects. Using more realistic
hydrodynamic models LN83 and LN85 derived approximate relations that connect the explosion energy ($E$),
the mass of the envelope ($M$), and the progenitor radius ($R_0$) to three observable quantities, namely,
the duration of the plateau, the absolute $V$ magnitude, and the photospheric
velocity observed in the middle of the plateau. These equations provide a simple and quick method
to derive $E$, $M$, and $R_0$ from observations of SNe~II-P without having to craft specific
models for each SN. A generalized analytic solution was subsequently worked out by \citet{popov93},
which proved in good agreement with the theoretical relations of \citet{litvinova85}.
So far these methods have been only applied to the one object (SN~1969L) which had sufficient observations
for this analysis. In this section I revisit this issue based on a larger sample of SNe~II-P, 
with the purpose to better understand the nature of such objects.

\begin{figure}
\epsscale{1.0}
\plotone{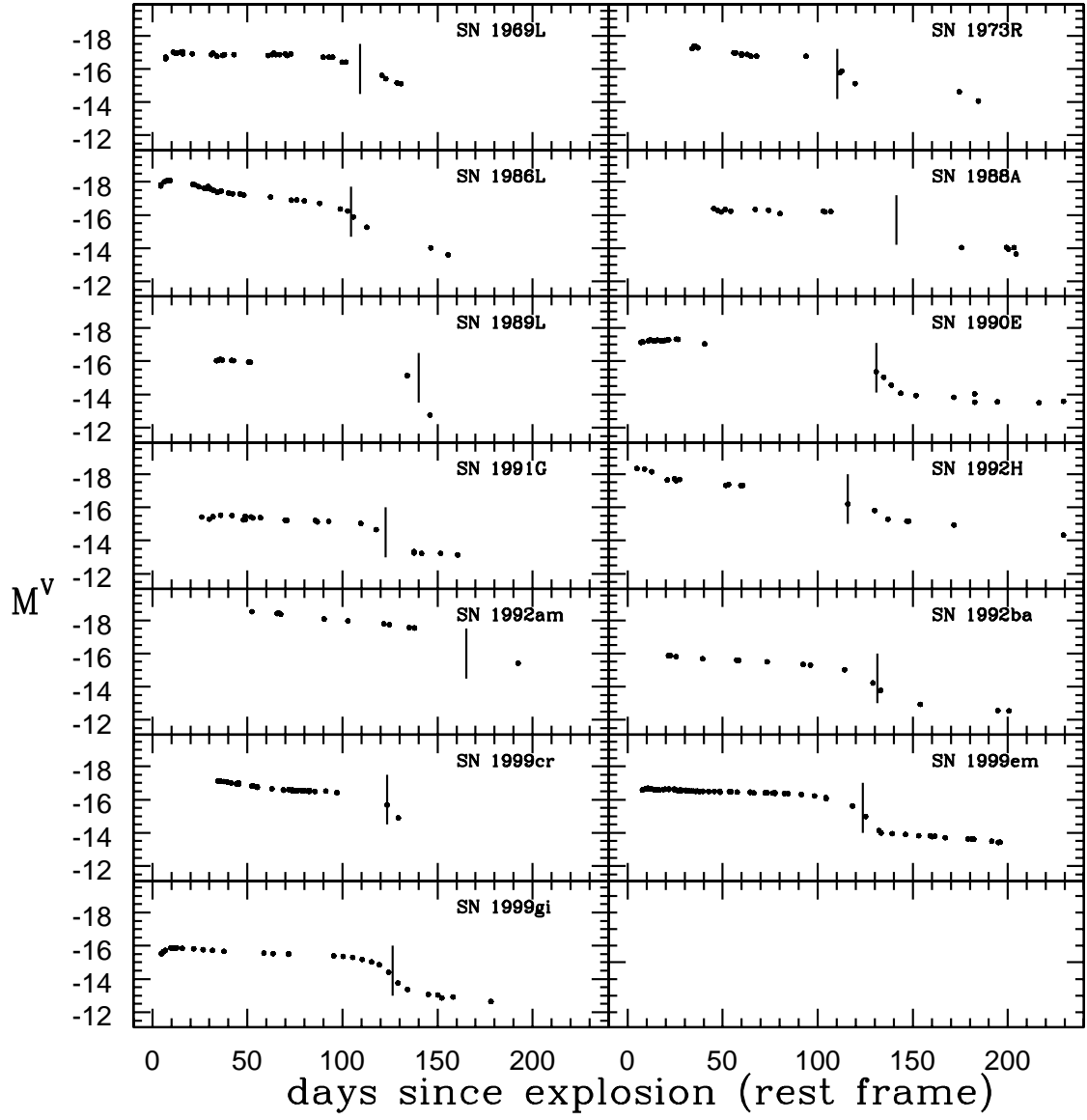}
\caption{Extinction corrected absolute $V$-band lightcurves of the 13 plateau SNe~II.
The vertical bars indicate the end of the plateau phase for each supernova.
\label{MLC_fig}}
\end{figure}

Of all the SNe~II-P listed in Table \ref{50t_tab} only 13 have sufficient data to apply
the method of LN85. In the top section of Table \ref{EMR_tab} I list such objects and the
observed quantities required for the LN85 analysis, in the following order: 
the time of explosion ($t_0$); the end of the plateau phase ($t_p$) defined as the time  
when the SN magnitude is near the mid-point between the plateau magnitude and that of the onset of the nebular phase;
the plateau visual magnitude ($V_p$) measured at $t=(t_0+t_p$)/2 (the middle of the plateau);
and the SN ejecta velocity at the middle of the plateau ($v_p$) measured from
the minimum of the Fe II~$\lambda$5169 line \footnote{Note that $V_p$ and $v_p$ are close
but not identical to the parameters $V_{50}$ and $v_{50}$ used in the section \ref{OP}.}.
Figure \ref{MLC_fig} shows the extinction-corrected absolute $V$-band lightcurves for 
the 13 SNe~II-P and the end of the plateau phase for each SN.

With these data and the formulas given by LN85 I can solve now for $E$, $M$, and $R_0$.
I attach 1-$\sigma$ uncertainties to each of the parameters from Monte Carlo simulations in which
I randomly vary the observed quantities according to the observational errors.
The resulting parameters are summarized in Table \ref{EMR_tab}.
Also included in Table \ref{EMR_tab} is SN~1987A which was modeled in detail by \citet{arnett96}.
Although SN~1987A showed an atypical lightcurve due to the compact nature of its blue supergiant progenitor,
it was not fundamentally different than ordinary SNe~II-P in the sense that it had
a hydrogen-rich envelope at the time of explosion. For this reason I include it in this
analysis. Also given in Table \ref{EMR_tab} are SN~1997D and SN~1999br, two low-luminosity
SNe~II-P recently modeled by \citet{zampieri02}. To my knowledge these are the
only 16 SNe with available physical parameters.

Among this sample, 9 SNe have explosion energies close to the canonical 1 foe value 
(1 foe=10$^{51}$ ergs), 6 objects exceed 2 foes, and one has only 0.6 foes.
SN~1992am and SN~1999br show the highest and lowest energies with 5.5
and 0.6 foes, respectively. This reveals that SNe~II encompass a wide range in explosion energies.
The ejected masses vary between 14 and 56 $M_\odot$. Although the uncertainties are large it is interesting
to note that, while stars born with more than 8 $M_\odot$ can in principle undergo
core collapse, they do not show up as SNe~II-P. Perhaps they undergo significant
mass loss before explosion and are observed as SNe~II-n or SNe~Ib/c. It proves
interesting also that stars as massive as 50 $M_\odot$ seem able to retain a significant fraction
of their H envelope and explode as SNe~II. Objects with $M$$>$35$~M_\odot$ are
supposed to lose their H envelope due to strong winds, and become Wolf-Rayet
stars before exploding \citep{woosley93}. This result suggests that
stellar winds in massive stars are not so strong as previously thought, perhaps due to
smaller metallicities. Except for four objects, the initial radii vary between
114 and 586 $R_\odot$. Within the error bars these values correspond to those measured for K and M red supergiants \citep{vanbelle99},
which lends support to the generally accepted view that the progenitors of SNe~II-P have extended atmospheres at
the time of explosion \citep{arnett96}. Three of the SNe~II-P of this sample, however, have
$R_0$$\sim$80 $R_\odot$ which corresponds to that of G supergiants.
This is somewhat odd because such objects are not supposed to have
plateau lightcurves but, instead, one like that of SN~1987A.
Note, however, that the uncertainties are quite large and it is possible that
these objects did explode as red supergiants.

\begin{figure}
\epsscale{1.0}
\plotone{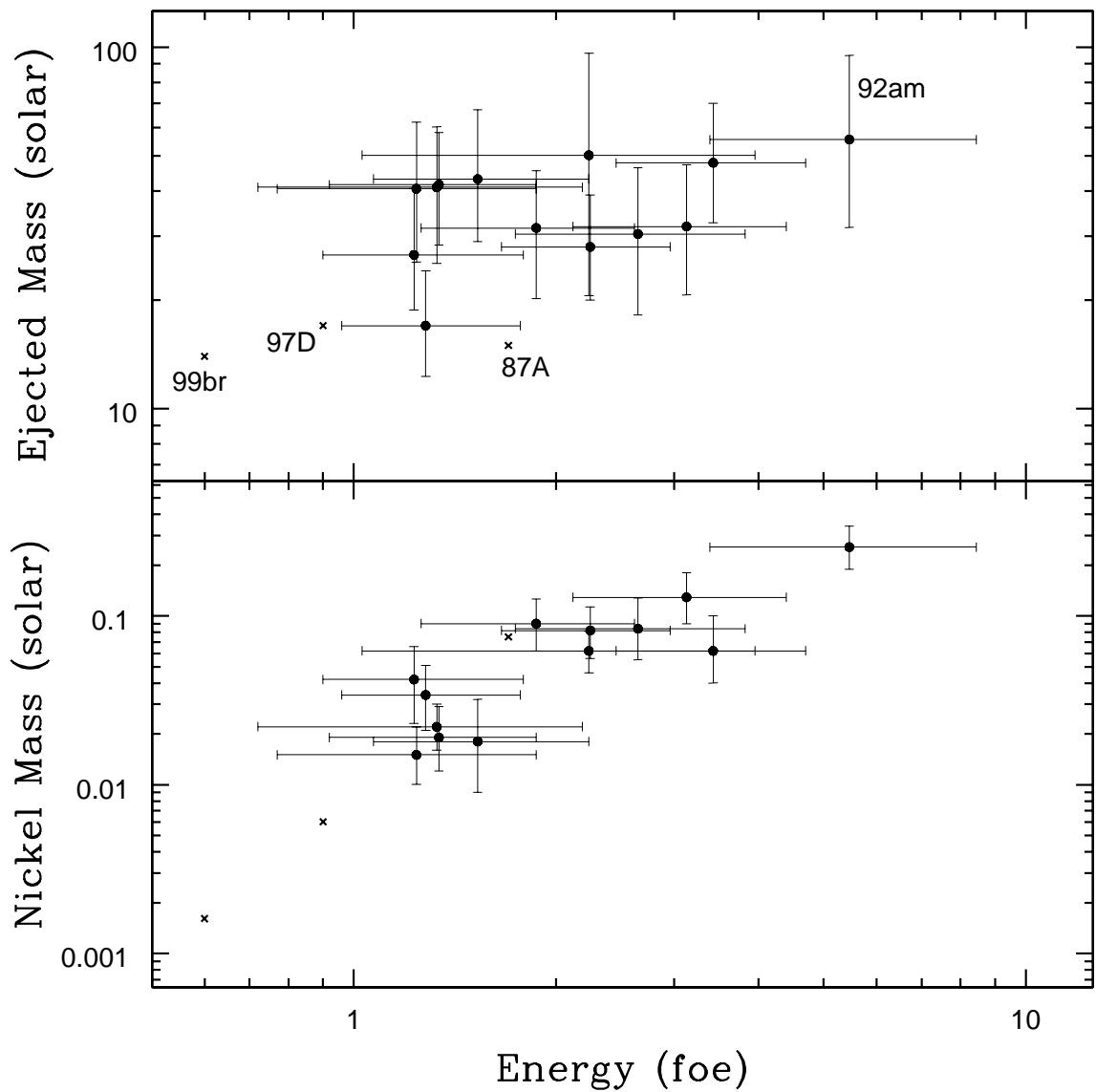}
\caption{Envelope mass and nickel mass of SNe~II, as a function of explosion
energy. Solid points represent the 13 SNe~II-P for which I was able to apply
the technique of \citet{litvinova85}. The three crosses correspond to SN~1987A,
SN~1997D, and SN~1999br which have been modeled in detail by \citet{arnett96}
and \citet{zampieri02}. The nickel yield for SN~1999br comes from this paper
(Table \ref{50t_tab}).
\label{ME_fig}}
\end{figure}

Figure \ref{ME_fig} shows $M$ and $M_{Ni}$ as a function of $E$ for the 16 SNe~II-P.
Despite the large error bars, this figure reveals that a couple of correlations
emerge from this analysis. The first interesting result (top panel) is that the
explosion energy appears to be correlated with the envelope mass, in the sense
that more massive progenitors produce higher energy SNe. This suggests that the outcome
of the core collapse is somehow determined by the mass of the envelope.  The second remarkable
result (bottom panel) is that SNe with greater energies produce more nickel (a result already
anticipated in section \ref{OP}, and previously suggested by \citet{blanton95}).
This could mean that greater temperatures and more nuclear burning are reached in such SNe,
and/or that less mass falls back onto the neutron star/black hole in more energetic explosions.

Before leaving this section it is necessary to mention some caveats about these results:

\noindent $\bullet$ The LN85 formulas were obtained from models with progenitor
masses and explosion energies below 2.9 $M_\odot$ and 16 foes, respectively.
Clearly some of the SNe~II-P in Table \ref{EMR_tab} lie outside the parameter
space explored by LN85 and my results involve extrapolating their formulas.
It will be necessary to expand the models to greater masses and energies
before we can truly believe that SNe~II-P have energies above 3 foes, progenitors with
$M$$\sim$50 $M_\odot$, and the correlations shown above.

\noindent $\bullet$ LN85 assumed that the plateau luminosity is fully powered by
shock-deposited energy and they neglected a contribution by the $^{56}$Co decay.
It will be interesting to generalize the models in order to find out how 
the radioactive heating and the distribution of $^{56}$Co affect the
results presented here.

\noindent $\bullet$ In the LN85 models the plateau phase is preceded by a brief transient
which lasts a few days, and the length of the plateau is measured from
the time at which this short phase ends. The data for the 13 SNe do
not show such transient, most likely because it is too short,
so I am forced to use the time of explosion for the beginning of the plateau.
This should lead to an overestimate ($\sim$2\%) of its length and
a small bias in the derived physical parameters. Given the difficulty
to measure the transient it would be desirable to re-derive the LN85
calibrations using a more operational definition of the onset of the plateau
such as the explosion time.

\noindent $\bullet$ The LN85 formulas use the velocity of the photosphere
($\tau$=2/3) as one of the input parameters. In my case I measure velocities
from the minimum of the Fe II~$\lambda$5169 line which is expected to arise
just above the thermalization surface (the region where the radiation field
forms). Since SNe~II have electron scattering dominated atmospheres, the
radiation field thermalizes well below the photosphere \citep{eastman96}
so that Fe II~$\lambda$5169 should underestimate the photospheric velocity.
In my thesis \citep{hamuy01a} I examined this by measuring true photospheric velocities
by cross-correlating (CC) the SN spectra with the Eastman et al. models.
This study showed systematic differences between the Fe and CC velocities
for individual objects but, curiously, no significant difference for the ensemble of SNe.
This suggests that on average the Fe method is a good estimator of the photospheric velocity,
although it may not work so well on an individual basis.

\noindent $\bullet$ To transform bolometric luminosities into $V$ magnitudes
LN65 employed bolometric corrections assuming that SNe~II have blackbody spectra.
SNe~II are not perfect blackbodies, of course. Using the theoretical spectra of
Eastman et al. I find that the bolometric corrections derived from
Planck functions are $\sim$0.2 mag too large for $T_{eff}$$\geq$6500 K, 
about right ($\pm$0.1 mag) between 5000$\leq$$T_{eff}$$\leq$6500 K,
and systematically low for $T_{eff}$$\leq$5000 K. It would be convenient
that the LN65 formulas were re-derived with improved corrections. 

\section{PROPERTIES OF CORE COLLAPSE SUPERNOVAE}
\label{PCCS}

Core collapse SNe can also be hosted by massive stars which have lost most or all
of their hydrogen-rich envelopes (SNe~Ib), and even most or all of their
helium envelopes (SNe~Ic). It proves interesting therefore to compare the physical
properties of such objects with those derived from SNe~II-P. A bibliographic search reveals
that there are only a handful well-studied SNe~Ib/c. Table \ref{SNIbc_tab} lists
such objects and the corresponding references from which their physical
parameters were obtained. 

In general, SNe~Ib/c have bell-shaped lightcurves with a rise time of $\sim$15-20 days, 
a fast-decline phase of $\sim$30 days, and a slower decline phase at a rate between 0.01-0.03 mag day$^{-1}$.
Unlike SNe~II-P the lightcurves of SNe~Ib/c are promptly powered by $^{56}$Ni $\rightarrow$ $^{56}$Co $\rightarrow$ $^{56}$Fe.
While the peak is determined by the amount of nickel synthesized in the explosion,
the width depends on the ability of the photons to diffuse out from the SN interior, which
is determined by the envelope mass and expansion velocity.
The early-time lightcurve, therefore, provides useful constraints on the $^{56}$Ni mass,
envelope mass, and kinetic energy \citep{arnett96}. Additional constraints on the kinetic energy
come from the Doppler broadening of the spectral lines. The late-time decline rate
reveals that a fraction of the $\gamma$-rays from the radioactive decay escape from
the SN ejecta without being thermalized and, therefore, can be used to quantify
the degree of $^{56}$Ni mixing in the SN interior. Nomoto and collaborators have
modeled SNe~Ib as helium stars that lose their hydrogen envelopes by mass transfer
to a binary companion, and SNe~Ic as C/O bare cores that lose their He envelope
in a second stage of mass transfer. In both cases they assume spherically
symmetric explosions. Table \ref{SNIbc_tab} summarizes the parameters
derived from such models for the 7 SNe~Ib/c.

\begin{figure}
\epsscale{1.0}
\plotone{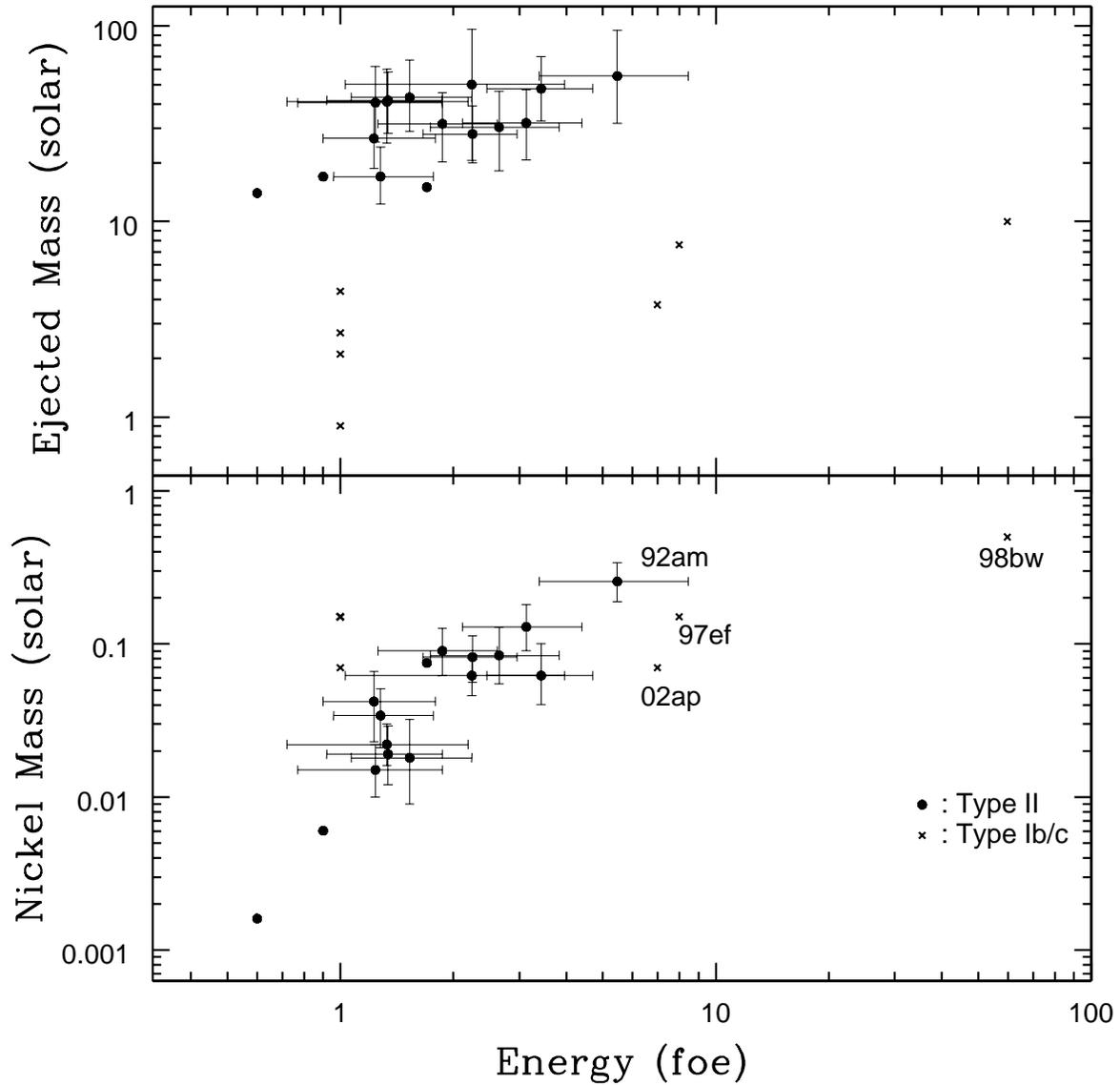}
\caption{Envelope mass and nickel mass of core collapse SNe, as a function of explosion
energy. Solid points are the same 16 SNe~II shown in Figure \ref{ME_fig}, and
crosses correspond to the 7 Type Ib/c SNe listed in Table \ref{SNIbc_tab}.
\label{ME_1_fig}}
\end{figure}

Figure \ref{ME_1_fig} shows envelope masses and nickel masses as a function of explosion
energy for the seven SNe~Ib/c along with the 16 SNe~II shown in Figure \ref{ME_fig}.
The top panel reveals that SNe~Ib/c appear to follow the same pattern shown by SNe~II,
namely, that SNe with greater envelope masses produce more energetic explosions.
The main difference between both subtypes, of course, is the vertical offset caused by
the strong mass loss suffered by SNe~Ib/c prior to explosion. From the bottom panel 
it is possible to appreciate that SN~1998bw was quite remarkable in explosion energy
(60 foes) and nickel mass (0.5 $M_\odot$) compared to all of the other core collapse
SNe. Owing to its extreme energy this object has been called hypernova. SN~1998bw
is also remarkable because it was discovered at nearly the same place and time as GRB 980425 \citep{galama98}.
The Type Ic SN~1997ef and SN~2002ap are located far below SN~1998bw in the energy scale
(8 and 7 foes, respectively), yet far above the normal SN~1994I. 
Despite their greater than normal energies, neither of these objects produced unusually higher nickel masses
compared to lower energy SNe~Ib/c.  Although the statistics are poor, it proves interesting that 
both SNe~Ib/c and SNe~II share the same location in this plane,
which suggests that the core physics of both subtypes may not be fundamentally different.

When the whole sample of SNe~II and SNe~Ib/c is considered it seems that there is
a continous distribution of energies below 8 foes. Within this regime it
appears that SNe~II can reach explosion energies comparable to that of the
Type Ib/c SN~1997ef and SN~2002ap. Although the definition of hypernova is
ambiguous, if SN~1997ef and SN~2002ap are included in this category \citep{nomoto00,mazzali02},
then at least one SN~II (1992am) also qualifies as a hypernova. Whether the
energy distribution is continuous above 8 foes remains to be seen when
more data become available. This will permit us to
understand if SN~1998bw belongs to a separate class of object or if it just lies
at the extreme of the family of core collapse SNe. At the moment it is fair to
say that there is only one firm supernovae/GRB association, and this object was
clearly exceptional regarding energy and nickel production within the SN context.

\section{CONCLUSIONS}
\label{CC}

I assembled photometric and spectroscopic data for 24 SNe~II-P which
allowed me to draw the following conclusions,

\noindent 1) As previously known, I recovered the result that SNe~II-P encompass
a wide range of $\sim$5 mag in plateau luminosities and a five-fold range in
expansion velocities. I recovered the luminosity-velocity relation previously
reported by \citet{hamuy02} which supports the claim that SNe~II-P have
a potential utility as cosmological probes. This empirical relation is
also supported by the theoretical models of LN83 and LN85.

\noindent 2) SNe~II-P encompass a factor of 10 in nickel masses between 0.0016 (SN~1999br)
and 0.26 $M_\odot$ (SN~1992am). There is clear evidence for a correlation in
the sense that SNe with brighter plateaus and greater expansion velocities
produce more nickel.

\noindent 3) There is a continuum in the properties of SNe~II-P from faint,
low-velocity, nickel-poor events such as SN~1997D and SN~1999br, and
bright, high-velocity, nickel-rich objects like SN~1992am. The correlations
between plateau luminosities, expansion velocities, and nickel masses
suggest that SNe~II-P constitute a one parameter family.

\noindent 4) Using the theoretical models of LN83 and LN85 I derived physical
parameters for a subset of 13 SNe. Including SN~1987A, SN~1997D, and SN~1999br
from previous studies I found that the explosion energies vary between
0.6 (SN~1999br) and 5.5 foes (SN~1992am), the ejected masses encompass
the range 14-56 $M_\odot$, and the progenitors' radii go from 80 to
600 $R_\odot$.

\noindent 5) Despite the large error bars, a couple of correlations emerge
from the previous analysis: (1) more massive progenitors produce more energetic
explosions, which suggests that the outcome of the core collapse is somewhat
determined by the envelope mass; (2) SNe with greater energies produce more
nickel.  Similar relationships appear to hold for Type Ib/c SNe, which suggests
that both Type II and Type Ib/c SNe share the same core physics.

\acknowledgments

\noindent
I am grateful to Brian Schmidt for his thorough referee report, 
to John Tonry for sending me his code with the peculiar parametric flow model,
and to Andrew McFayden for useful discussions during the preparation of this paper.
I thank the Lorentz Center at Leiden University where I was able to complete a first
draft of this paper.
Support for this work was provided by NASA through Hubble Fellowship grant HST-HF-01139.01-A
awarded by the Space Telescope Science Institute, which is operated by the Association
of Universities for Research in Astronomy, Inc., for NASA, under contract NAS 5-26555.
This research has made use of the NASA/IPAC Extragalactic Database (NED), which is operated by the
Jet Propulsion Laboratory, California Institute of Technology, under
contract with the National Aeronautics and Space Administration.
This research has made use of the SIMBAD database, operated at CDS, Strasbourg, France.

\clearpage

\begin{deluxetable} {ccccccccc}
\rotate
\tabletypesize{\scriptsize}
\tablecolumns{9}
\tablenum{1}
\tablewidth{0pc}
\tablecaption{General Data for Type II Supernovae \label{SNII_tab}}
\tablehead{
\colhead{SN} &
\colhead{$cz_{helio}$} &
\colhead{Redshift} &
\colhead{$A_{GAL}(V)$} &
\colhead{$A_{host}(V)$} &
\colhead{$A_{host}(V)$} &
\colhead{$A_{host}(V)$} &
\colhead{Distance} &
\colhead{Distance} \\
\colhead{} &
\colhead{($km~s^{-1}$)} &
\colhead{Source} &
\colhead{$\pm$0.06} &
\colhead{$B-V$} &
\colhead{$V-I$} &
\colhead{$\pm$0.3} &
\colhead{($Mpc$)} &
\colhead{Method} }
\startdata

1968L  &   516  & 2 & 0.219 & -0.90 &\nodata & 0.00 &   4.1(1.0) & SBF (Cen A Group) \\
1969L  &   518  & 2 & 0.205 & -0.70 &\nodata & 0.00 &  10.0(1.0) & SBF (N1023 Group) \\
1970G  &   241  & 2 & 0.028 & -1.20 &\nodata & 0.00 &   7.4(0.3) & Cepheids \\
1973R  &   727  & 2 & 0.107 &  1.40 &\nodata & 1.40 &  10.3(0.8) & Cepheids \\
1986I  &  2407  & 2 & 0.129 &\nodata&  0.20  & 0.20 &  17.0(1.0) & SBF (Virgo Group) \\
1986L  &  1292  & 2 & 0.099 &  0.30 &\nodata & 0.30 &  18.7(2.4) & SBF model \\
1988A  &  1519  & 2 & 0.136 & -0.40 &\nodata & 0.00 &  17.0(1.0) & SBF (Virgo Group) \\
1989L  &  1313  & 2 & 0.123 & -0.60 &  0.90  & 0.15 &  17.0(2.4) & SBF model \\         
1990E  &  1241  & 2 & 0.082 &  1.00 &  1.90  & 1.45 &  18.2(2.4) & SBF model \\
1990K  &  1584  & 2 & 0.047 &  0.05 &  0.35  & 0.20 &  23.2(2.4) & SBF model \\
1991al &  4572  & 1 & 0.168 & -0.30 &  0.10  & 0.00 &  65.9(2.4) & Redshift \\
1991G  &   757  & 2 & 0.065 &\nodata&  0.00  & 0.00 &  14.7(1.0) & SBF (U Ma Group) \\
1992H  &  1793  & 2 & 0.054 &  0.00 &\nodata & 0.00 &  29.4(2.4) & SBF model \\ 
1992af &  5611  & 1 & 0.171 & -0.40 & -0.20  & 0.00 &  80.0(2.4) & Redshift \\
1992am & 14310  & 2 & 0.164 &  0.35 &  0.20  & 0.28 &  206.0(2.4)& Redshift \\
1992ba &  1104  & 2 & 0.193 & -0.15 &  0.15  & 0.00 &  15.2(2.4) & SBF model \\         
1993A  &  8790  & 1 & 0.572 &  0.00 &  0.10  & 0.05 &  131.4(2.4)& Redshift \\
1993S  &  9896  & 2 & 0.054 &  1.00 &  0.40  & 0.70 &  141.9(2.4)& Redshift \\
1999br &   969  & 2 & 0.078 &  0.50 &  0.80  & 0.65 &  10.8(2.4) & SBF model \\ 
1999ca &   2791 & 2 & 0.361 &  0.85 &  0.50  & 0.68 &  45.7(2.4) & Redshift \\
1999cr &  6069  & 1 & 0.324 & -0.75 &  0.10  & 0.00 &  93.8(2.4) & Redshift  \\
1999eg &  6703  & 2 & 0.388 & -0.15 &  0.05  & 0.00 &  95.5(2.4) & Redshift  \\
1999em &   717  & 2 & 0.130 &  0.18 &  0.18  & 0.18 &  10.7(2.4) & SBF model \\
1999gi &   592  & 2 & 0.055 &  0.50 &  0.85  & 0.68 &   9.0(2.4) & SBF model \\

\enddata
\tablerefs{
(1) \citet{hamuy01a};
(2) NASA/IPAC Extragalactic Database.}
\end{deluxetable}

\clearpage

\begin{deluxetable} {ccccccccccc}
\rotate
\tabletypesize{\scriptsize}
\tablecolumns{11}
\tablenum{2}
\tablewidth{0pc}
\tablecaption{Observed Parameters for Type II Supernovae \label{50t_tab}}
\tablehead{
\colhead{SN} &
\colhead{$t_0$} & 
\colhead{$V_{50}$} &
\colhead{$M^V_{50}$} &
\colhead{$v_{50}$} &
\colhead{$t_{50}$} &
\colhead{$V_t$} &
\colhead{$t_t$} & 
\colhead{$M_{Ni}$} &
\colhead{Phot.} &
\colhead{Spec.} \\
\colhead{} &
\colhead{(JD - 2,400,000)} &
\colhead{} &
\colhead{} &
\colhead{($\pm300~km~s^{-1}$)} &
\colhead{(JD - 2,400,000)} &
\colhead{} &
\colhead{(JD - 2,400,000)} &
\colhead{($M_\odot$)} &
\colhead{Source} &
\colhead{Source} }
\startdata

1968L  & 40039.5(5) &  12.03(08)  & -16.25(55)   &    4020     &   40089.6 & \nodata   & \nodata  & \nodata                            & 1        &  1 \\
1969L  & 40550.5(5) &  13.35(06)  & -16.85(37)   &    4841     &   40600.6 & 17.16(10) & 40860.0  & 0.082$_{\rm -0.026}^{+0.034}$      & 2        &  2,3 \\
1970G  & 40768.5(30)&  12.10(15)  & -17.27(35)   &    5041     &   40818.5 & 16.25(10) & 40979.5  & 0.037$_{\rm -0.012}^{+0.019}$      & 4,5      &  3,6 \\
1973R  & 42008.5(15)&  14.56(05)  & -17.01(35)   &    5092     &   42058.6 & 17.23(28) & 42187.8  & 0.084$_{\rm -0.030}^{+0.044}$      & 7        &  7   \\
1986I  & 46563.3(4) &  14.55(20)  & -16.93(34)   &    3623     &   46613.7 & 16.93(06) & 46758.0  & 0.117$_{\rm -0.031}^{+0.039}$      & 8,9      &  8  \\
1986L  & 46707.9(4) &  14.57(05)  & -17.19(41)   &    4150     &   46758.1 & 18.16(15) & 46864.1  & 0.034$_{\rm -0.011}^{+0.018}$      & 10       &  10 \\
1987A  & \nodata    &  \nodata    & \nodata      &   \nodata   &   \nodata & \nodata   & \nodata  & 0.075$^{a}$                        & \nodata  & \nodata \\
1988A  & 47163.0(7) &  15.00(05)  & -16.29(34)   &    4613     &   47213.3 & 19.11(24) & 47531.6  & 0.062$_{\rm -0.020}^{+0.029}$      & 11,12,13 &  13,14 \\
1989L  & 47650.0(15)&  15.47(05)  & -15.96(43)   &    3529     &   47700.2 & 18.67(12) & 47796.7  & 0.015$_{\rm -0.005}^{+0.008}$      & 15       &  16 \\
1990E  & 47932.6(5) &  15.90(20)  & -16.93(43)   &    5324     &   47982.8 & 19.59(25) & 48191.2  & 0.062$_{\rm -0.022}^{+0.031}$      & 17,18    &  17,14 \\
1990K  & 47970.0(30)&  14.50(20)  & -17.57(45)   &    6142     &   48020.3 & 18.84(07) & 48178.6  & 0.039$_{\rm -0.014}^{+0.022}$      & 19,20    &  14 \\
1991al & 48410.0(30)&  16.62(05)  & -17.64(34)   &    7330     &   48460.8 & 19.44(08) & 48555.5  & 0.095$_{\rm -0.028}^{+0.048}$      & 14       &  14 \\
1991G  & 48280.0(5) &  15.53(07)  & -15.37(33)   &    3347     &   48330.1 & 17.69(03) & 48428.0  & 0.022$_{\rm -0.006}^{+0.008}$      & 21       &  21 \\
1992H  & 48661.0(10)&  14.99(04)  & -17.41(36)   &    5463     &   48711.3 & 18.58(06) & 48943.2  & 0.129$_{\rm -0.037}^{+0.053}$      & 22,23    &  23 \\
1992af & 48736.0(30)&  17.06(20)  & -17.63(37)   &    5322     &   48786.9 & 19.42(03) & 48891.6  & 0.156$_{\rm -0.051}^{+0.078}$      & 14       &  14 \\
1992am & 48778.1(11)&  18.44(05)  & -18.57(31)   &    7868     &   48830.5 & 21.60(10) & 48979.7  & 0.256$_{\rm -0.070}^{+0.099}$      & 14       &  14 \\
1992ba & 48883.2(5) &  15.43(05)  & -15.67(43)   &    3523     &   48933.4 & 18.55(05) & 49081.6  & 0.019$_{\rm -0.007}^{+0.009}$      & 14       &  14 \\
1993A  & 48985.5(10)&  19.64(05)  & -16.57(33)   &    4290     &   49037.0 & \nodata   & \nodata  & \nodata                            & 14       &  14 \\
1993S  & 49130.0(10)&  18.96(05)  & -17.55(30)   &    4569     &   49181.6 & \nodata   & \nodata  & \nodata                            & 14       &  14 \\
1997D  & \nodata    &  \nodata    & \nodata      &   \nodata   &   \nodata & \nodata   & \nodata  & 0.006$^{b}$                        & \nodata  & \nodata \\
1999br & 51277.9(3) &  17.58(05)  & -13.32(53)   &    1545     &   51328.1 & 22.68(08) & 51643.7  & 0.0016$_{\rm -0.0008}^{+0.0011}$   & 14,24    &  14 \\
1999ca & 51280.0(10)&  16.65(05)  & -17.69(35)   &    5353     &   51330.5 & 21.10(10) & 51484.8  & 0.038$_{\rm -0.010}^{+0.017}$      & 14       &  14 \\
1999cr & 51221.5(10)&  18.33(05)  & -16.86(29)   &    4389     &   51272.5 & 20.29(20) & 51353.5  & 0.090$_{\rm -0.027}^{+0.034}$      & 14       &  14 \\
1999eg & 51437.2(8) &  18.65(05)  & -16.64(31)   &    4012     &   51488.3 & \nodata   & \nodata  & \nodata                            & 14       &  14 \\
1999em & 51474.0(3) &  13.98(05)  & -16.48(52)   &    3757     &   51524.1 & 16.70(03) & 51636.8  & 0.042$_{\rm -0.019}^{+0.027}$      & 25,26    &  25,27 \\
1999gi & 51518.2(3) &  14.91(05)  & -15.60(58)   &    3617     &   51568.3 & 17.62(05) & 51676.6  & 0.018$_{\rm -0.009}^{+0.013}$      & 28       &  28 \\

\enddata
\tablenotetext{a} {From \citet{arnett96}}
\tablenotetext{b} {From \citet{zampieri02}}
\tablerefs{
(1) \citet{wood74};
(2) \citet{ciatti71};
(3) \citet{kirshner74};
(4) \citet{winzer74};
(5) \citet{barbon73};
(6) \citet{pronik76};
(7) \citet{ciatti77};
(8) \citet{pennypacker89};
(9) \citet{tsvetkov88};
(10) M. M. Phillips \& S. Kirhakos (2000) (private communication);
(11) \citet{ruiz90};
(12) \citet{benetti91};
(13) \citet{turatto93};
(14) \citet{hamuy01a};
(15) B. Schmidt (2002) (private communication);
(16) \citet{schmidt94a};
(17) \citet{schmidt93};
(18) \citet{benetti94};
(19) \citet{cappellaro95};
(20) M. M. Phillips (2000) (private communication);
(21) \citet{blanton95};
(22) \citet{tsvetkov94};
(23) \citet{clocchiatti96};
(24) A. Pastorello \& L. Zampieri (2002) (private communication);
(25) \citet{hamuy01b};
(26) N. B. Suntzeff (private communication);
(27) \citet{leonard02a};
(28) \citet{leonard02b}
.}
\end{deluxetable}

\clearpage

\begin{deluxetable} {ccccccccc}
\rotate
\tabletypesize{\scriptsize}
\tablecolumns{9}
\tablenum{3}
\tablewidth{0pc}
\tablecaption{Observed and Physical Parameters for Type II Supernovae \label{EMR_tab}}
\tablehead{
\colhead{SN} &
\colhead{$t_0$} &
\colhead{$t_p$} &
\colhead{$V_p$} &
\colhead{$v_p$} &
\colhead{Energy} &
\colhead{Ejected Mass} &
\colhead{Initial Radius} &
\colhead{References} \\
\colhead{} &
\colhead{(JD - 2,400,000)} &
\colhead{(JD - 2,400,000)} &
\colhead{} &
\colhead{($\pm300~km~s^{-1}$)} &
\colhead{(10$^{51}$ ergs)} &
\colhead{($M_\odot$)} &
\colhead{($R_\odot$)} &
\colhead{}  }
\startdata

1969L  &  40550.5(5)  & 40660.0(7)  & 13.34(06) & 4562    & 2.3$_{\rm -0.6}^{+0.7}$  & 28$_{\rm  -8}^{+11}$ & 204$_{\rm  -88}^{+150}$ & 1 \\
1973R  &  42008.5(15) & 42119.0(7)  & 14.61(05) & 4823    & 2.7$_{\rm -0.9}^{+1.2}$  & 31$_{\rm -12}^{+16}$ & 197$_{\rm  -78}^{+128}$ & 1 \\
1986L  &  46707.9(4)  & 46813.0(7)  & 14.64(05) & 4037    & 1.3$_{\rm -0.3}^{+0.5}$  & 17$_{\rm  -5}^{ +7}$ & 417$_{\rm -193}^{+304}$ & 1 \\
1988A  &  47163.0(7)  & 47305.0(35) & 15.04(05) & 3537    & 2.2$_{\rm -1.2}^{+1.7}$  & 50$_{\rm -30}^{+46}$ & 138$_{\rm  -42}^{ +80}$ & 1 \\
1989L  &  47650.0(15) & 47790.7(7)  & 15.68(05) & 2800    & 1.2$_{\rm -0.5}^{+0.6}$  & 41$_{\rm -15}^{+22}$ & 136$_{\rm  -65}^{+118}$ & 1 \\
1990E  &  47932.6(5)  & 48063.9(10) & 16.00(20) & 4552    & 3.4$_{\rm -1.0}^{+1.3}$  & 48$_{\rm -15}^{+22}$ & 162$_{\rm  -78}^{+148}$ & 1 \\
1991G  &  48280.0(5)  & 48403.0(7)  & 15.61(07) & 3030    & 1.3$_{\rm -0.6}^{+0.9}$  & 41$_{\rm -16}^{+19}$ &  70$_{\rm  -31}^{ +73}$ & 1 \\
1992H  &  48661.0(10) & 48777.5(10) & 15.07(04) & 5084    & 3.1$_{\rm -1.0}^{+1.3}$  & 32$_{\rm -11}^{+16}$ & 261$_{\rm -103}^{+177}$ & 1 \\
1992am &  48778.1(11) & 48951.1(29) & 18.78(05) & 5097    & 5.5$_{\rm -2.1}^{+3.0}$  & 56$_{\rm -24}^{+40}$ & 586$_{\rm -212}^{+341}$ & 1 \\
1992ba &  48883.2(5)  & 49015.3(7)  & 15.56(05) & 2954    & 1.3$_{\rm -0.4}^{+0.5}$  & 42$_{\rm -13}^{+17}$ &  96$_{\rm  -45}^{+100}$ & 1 \\
1999cr &  51221.5(10) & 51347.5(10) & 18.50(05) & 3858    & 1.9$_{\rm -0.6}^{+0.8}$  & 32$_{\rm -12}^{+14}$ & 224$_{\rm  -81}^{+136}$ & 1 \\
1999em &  51474.0(3)  & 51598.0(5)  & 14.02(05) & 3290    & 1.2$_{\rm -0.3}^{+0.6}$  & 27$_{\rm  -8}^{+14}$ & 249$_{\rm -150}^{+243}$ & 1 \\
1999gi &  51474.0(3)  & 51645.0(5)  & 14.98(05) & 3168    & 1.5$_{\rm -0.5}^{+0.7}$  & 43$_{\rm -14}^{+24}$ &  81$_{\rm  -51}^{+110}$ & 1 \\
\cutinhead{Supernovae From Other Sources}
1987A  & \nodata      & \nodata     & \nodata   & \nodata & 1.7                      & 15                   & 42.8                    & 2 \\
1997D  & \nodata      & \nodata     & \nodata   & \nodata & 0.9                      & 17                   & 128.6                   & 3 \\
1999br & \nodata      & \nodata     & \nodata   & \nodata & 0.6                      & 14                   & 114.3                   & 3 \\

\enddata
\tablerefs{
(1) This paper;
(2) \citet{arnett96};
(3) \citet{zampieri02}
.}

\end{deluxetable}

\clearpage

\begin{deluxetable} {cccccc}
\tablecolumns{6}
\tablenum{4}
\tablewidth{0pc}
\tablecaption{Physical Parameters for Type Ib/c Supernovae \label{SNIbc_tab}}
\tablehead{
\colhead{SN} &
\colhead{Type} &
\colhead{Energy} &
\colhead{Ejected Mass} &
\colhead{Nickel Mass}  &
\colhead{References} \\
\colhead{} &
\colhead{} &
\colhead{(10$^{51}$ ergs)} &
\colhead{($M_\odot$)} &
\colhead{($M_\odot$)} &
\colhead{} }
\startdata

1983I   & Ic &  1.0 &  2.1  & 0.15 & 1  \\
1983N   & Ib &  1.0 &  2.7  & 0.15 & 1  \\
1984L   & Ib &  1.0 &  4.4  & 0.15 & 1  \\
1994I   & Ic &  1.0 &  0.9  & 0.07 & 2  \\
1997ef  & Ic &  8.0 &  7.6  & 0.15 & 2 \\
1998bw  & Ic & 60.0 & 10.0  & 0.50 & 2 \\
2002ap  & Ic &  7.0 &  3.75 & 0.07 & 3  \\

\enddata
\tablerefs{
(1) \citet{shigeyama90};
(2) \citet{nomoto00};
(3) \citet{mazzali02}
.}
\end{deluxetable}

\clearpage

\end{document}